\documentclass[aps,twocolumn]{revtex4}
\usepackage{graphicx}

\newcommand{\aos}{a_{\rm osc}}
\newcommand{\rtf}{R_{\mathrm{TF}}}
\newcommand{\beq}{\begin{equation}}
\newcommand{\enq}{\end{equation}}
\newcommand{\rr}{{\bf r}}
\newcommand{\ggp}{\Gamma_{\mathrm{GP}}}
\newcommand{\Omb}{\bar{\Omega}}
\newcommand{\tb}{\bar{t}}

\begin{document}

\title{Period-doubled breathing in trapped Bose-Einstein condensates}
\author{Emil Lundh}
\affiliation{Helsinki Institute of Physics, PL 64, FIN-00014
Helsingin yliopisto, Finland}
\affiliation{Department of Physics, Royal Institute of Technology,
Albanova, SE-106 91 Stockholm, Sweden\footnote{Present address.}}

\begin{abstract}
    The response of a trapped Bose-Einstein condensed gas to a 
    periodic driving force is studied theoretically in the framework 
    of the nonlinear Gross-Pitaevskii equation. The monopole 
    mode is driven by periodical modulation of the frequency of the 
    isotropic harmonic trapping potential. Period-doubled motion is 
    shown to occur when the driving is strong and close to resonance 
    with the monopole mode. 
    A variational model predicts chaotic oscillations but is found to 
    disagree with the full solutions to the Gross-Pitaevskii equation.
\end{abstract}
\maketitle

A trapped Bose-Einstein condensed gas is a damped, nonlinear 
oscillator. The modes of oscillation result from an interplay between 
dispersion, the repulsive inter-particle interactions and the confining trap 
potential \cite{stringari}, and the damping is offered by friction 
against thermal atoms \cite{jiladamping,theodamping}. Interesting physics can 
therefore be anticipated if the trap is modulated so as to drive the modes 
of oscillation: periodically forced, nonlinear oscillators can generally 
be expected to 
exhibit period-doubled motion and chaos for certain parameter values. 
Textbook examples include the Duffing oscillator and the forced, damped 
pendulum \cite{ott}.

In this paper, we shall study the oscillations of 
Bose-Einstein condensed gases subject to periodic driving forces, 
with particular attention to the conditions for period-doubled motion 
and the onset of chaos. 
The basis for all calculations is the 
Gross-Pitaevskii equation (GPE) for the condensate wave function 
$\psi(\rr,t)$,
\cite{gross,pitaevskii}
\begin{equation}\label{gpe}
  i\hbar\frac{\partial \psi}{\partial t} =
  -\frac{\hbar^{2}}{2m}\nabla^{2}\psi +
  V(\rr,t)\psi + 
  \frac{4\pi\hbar^{2}a}{m} |\psi|^{2}\psi.
\end{equation}
The wave function $\psi(\rr,t)$ determines the local fluid density and 
velocity through $n(\rr,t) = |\psi(\rr,t)|^2$ and 
$\mathbf{v}(\rr,t) = (\hbar/m) \nabla \arg \psi(\rr,t)$; it 
is normalized to the number of condensed atoms $N$.
The nonlinear term in the GPE 
describes inter-particle
interactions in the 
s-wave approximation and its strength is determined by the s-wave 
scattering length $a$.
The trap potential $V(\rr,t)$ is a 
harmonic-oscillator potential which in this study shall be taken to be 
isotropic and periodically modulated in time:
\beq
V(\rr,t) = \frac12m\omega^2
r^2\left(1+\Delta \cos \Omega t \right).
\enq
Assuming that the density profile stays fixed and only its overall 
radial extent is allowed to vary, the cloud size is described by an 
effective equation \cite{cd1},
\begin{equation}
  \ddot{R} = - R\left(1 + \Delta\cos\Omb \tb \right)
  + \frac{1}{R^{4}} - 2 \gamma \dot{R}.\label{req}
\end{equation} 
This equation is often derived in a variational scheme \cite{thebook}
and will 
therefore be referred to as ``variational'' in the following.
The first term on the right hand side represents the 
trap potential including the periodic modulation. The 
inverse-power term originates from the 
inter-particle interactions. The kinetic energy associated with 
the density gradient is assumed negligible in comparison with the 
trap and interaction energies and the corresponding term is 
therefore not included in Eq.\ (\ref{req}); 
this approximation holds when the 
scattering length is large and the mean density is high 
\cite{thebook}. 
The unit of length is taken to be the Thomas-Fermi radius 
$\rtf=\aos (15 Na/\aos)^{1/5}$, where 
$\aos = (\hbar/m \omega)^{1/2}$ is the oscillator length, and the 
time and frequency have been scaled by the trap frequency 
$\omega$, so that $\Omb = \Omega/\omega$ and $\tb=\omega t$. We 
shall hereafter work only in these dimensionless units, and we 
therefore immediately drop the bar on $t$ and $\Omega$. 
The final linear friction term in Eq.\ (\ref{req}) is 
introduced by hand and represents the 
dissipation due to the thermal gas. 
For the physical situation at hand, damping 
plays a pivotal role and it is therefore important to 
allow for this effect.
This phenomenological treatment of the dissipation 
is known to well reproduce the experimental findings 
of Ref.\ \cite{jiladamping}. The friction is provided by Landau 
damping and depends on temperature through the relation 
\cite{theodamping}
\begin{equation}
  \frac{\gamma}{\omega_m} = 
  0.645
  \frac{1}{N^{2/3}}
  \left(\frac{Na}{\aos}\right)^{4/5}
  \frac{T/T_c}{
    \left[1-\left(T/T_c\right)^{3}\right]^{1/5}},
  \label{eq:gammaT}
\end{equation}
where $\omega_m$ is the frequency of the mode in question, 
and $T_c$ is the critical Bose-Einstein condensation temperature. The 
damping coefficient $\gamma = 0.02$, which is used in much of 
the calculations of this paper, corresponds to a cloud of 
$N=10^{6}$ atoms with $\aos/a=100$ at a temperature $T=0.1T_{c}$.

The virtue of the variational equation (\ref{req}) is of course 
its simplicity, allowing for fast numerical solution, so that large 
parameter spaces can be covered, and in a few instances the existence 
of analytical solutions. However, the results obtained 
this way must of course be double-checked against the solution of
the full Gross-Pitaevskii equation, Eq.\ (\ref{gpe}), to find out  
whether deformation of the density profile plays a crucial 
role. For the present case of isotropic forcing, we shall assume that the 
cloud stays spherically symmetric. 
The Gross-Pitaevskii equation, thus made effectively one-dimensional, 
is propagated in time with the Crank-Nicolson method. 
Dissipation is represented by letting the time be complex, so 
that in solving the discretized equation, the time step is 
$\Delta t = \Delta t_R(1 + i\ggp)$. It is found numerically that the 
parameter $\ggp$ is related to the damping of the breathing mode 
through $\gamma \approx 3\ggp$.
    
The problem depends on three physical 
parameters: the amplitude $\Delta$ of the perturbation, the 
frequency $\Omega$ of the perturbation, and the 
damping $\gamma$ \cite{fotnot1}.
Luckily, the dynamics of the cloud is found not 
to depend sensitively on the latter, which is difficult to control 
or even measure with great precision. When the system is not 
driven, $\Delta=0$, the 
system performs damped sinusoidal 
oscillations with the frequency of the 
breathing mode, which is equal to 
$\sqrt{5}$ in the strong-coupling limit considered here. 
When the amplitude of the driving force $\Delta$ is 
nonzero, but not too large, we expect the motion to be frequency locked, 
i.\ e.\ it performs breathing motion that is 
periodic with the frequency $\Omega$, or possibly 
fractions thereof, in which case we have period-multiplied motion. 
When $\Delta >1$, the instantaneous trapping potential will for a 
certain time interval during each period not be confining, because 
the instantaneous squared trap frequency, 
$\omega^2 (1+\Delta\cos \Omega t)$, 
turns negative. This does not, however, necessarily imply that 
atoms are lost, because such events happen on a finite time scale: 
if the trap is opened but then sufficiently rapidly closed again, 
the atoms remain confined.
For large enough $\Delta$, the system does indeed turn unstable, 
but the threshold value
depends sensitively on $\Omega$ (and less sensitively on $\gamma$).
    
The region of instability of the variational Eq.\ (\ref{req}) have 
been examined in Refs.\ \cite{GarciaShort,GarciaLong}. For small forcing 
amplitudes $\Delta$, the instabilities occur in the vicinity of the 
frequency values $\Omega=2/n$, $n=1,2,3\ldots$, and as $\Delta$ is 
increased the instability regions grow larger. Since instability
is associated with large-amplitude motion, the relevant frequency is 
in fact the bare trap frequency (which is 1 in the units currently 
employed), rather than the quadrupole frequency $\sqrt{5}$, and the 
instability regime is the same as that for the 
Mathieu equation. The inclusion of damping somewhat shrinks the 
instability regimes. This analysis predicts an appreciable 
domain for stability even for values of $\Delta$ well above unity, 
especially when $\Omega$ is large. In fact, when $\Omega$ approaches 
infinity, the threshold value of $\Delta$ for instability also 
approaches infinity. 
    
\begin{figure}
    \includegraphics[width=0.9\columnwidth]{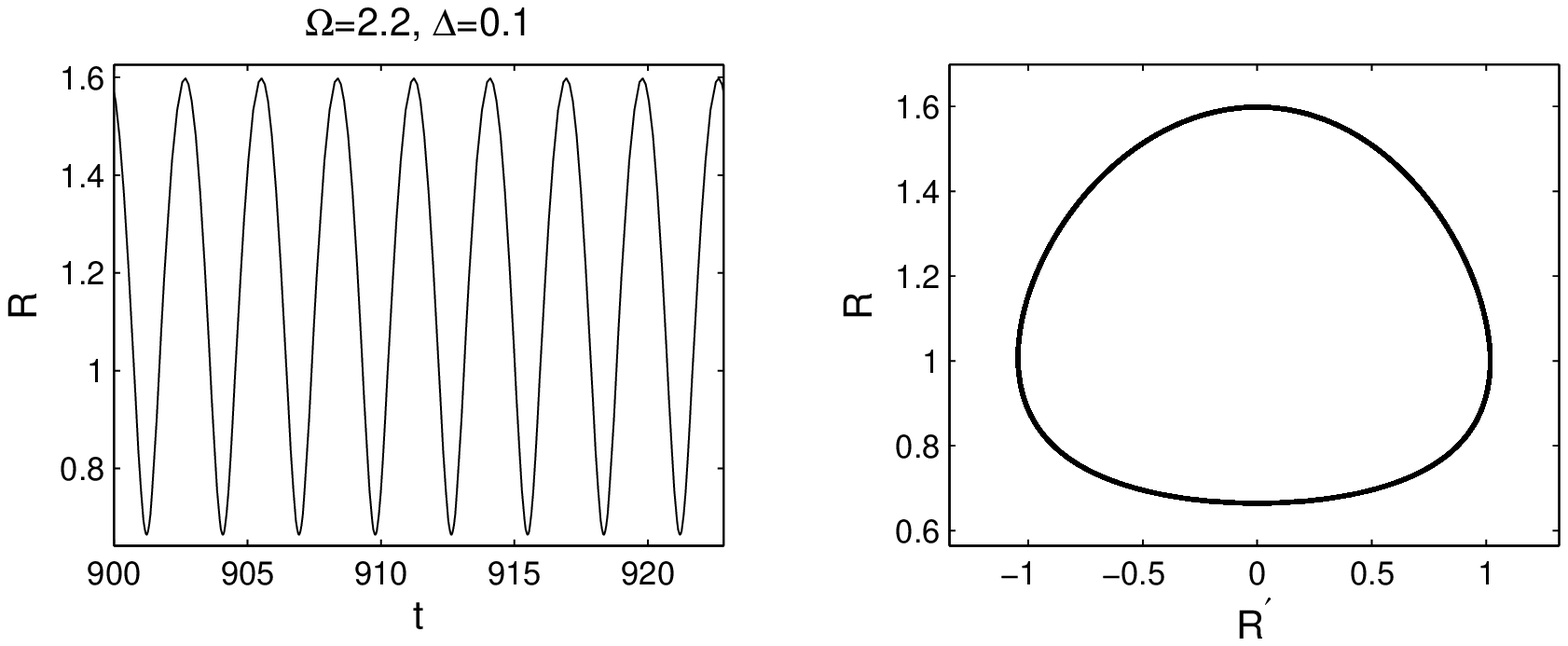}
    \includegraphics[width=0.9\columnwidth]{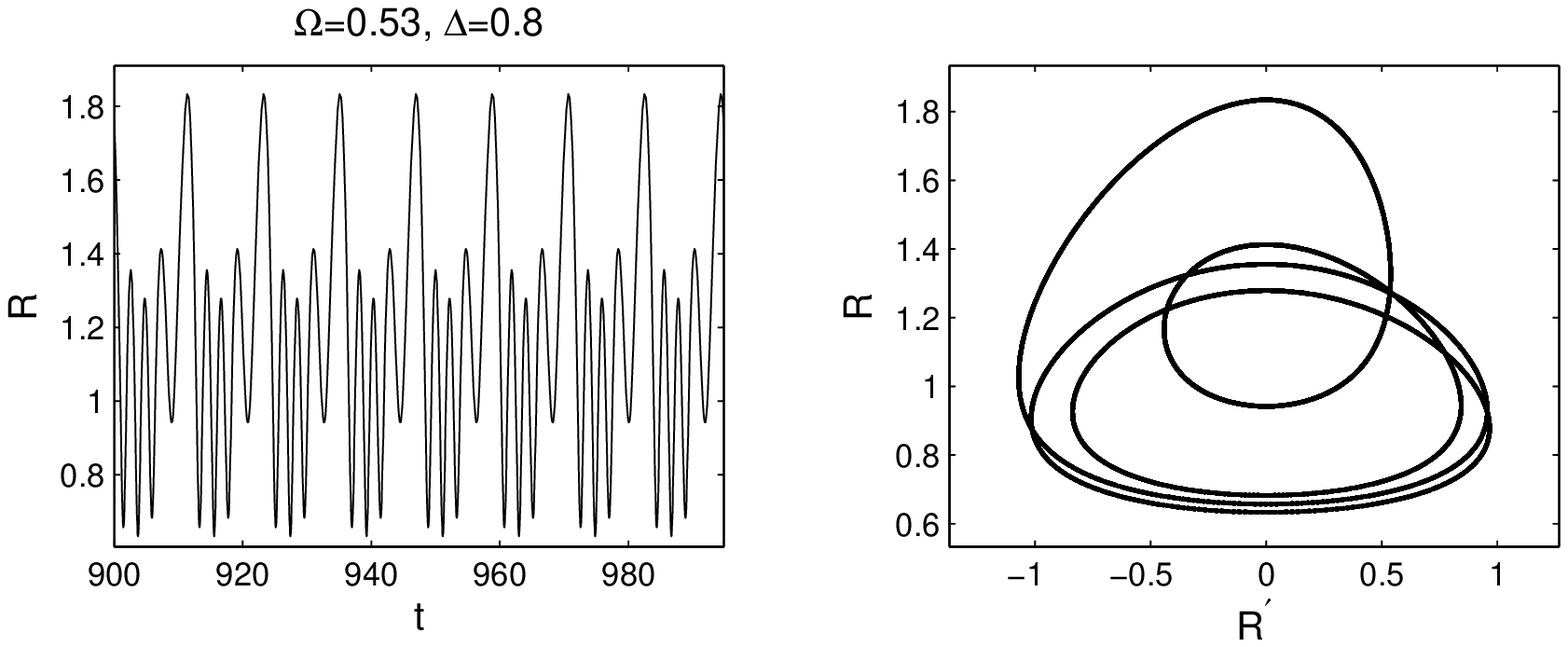}
    \caption{\label{fig:psplots} Time evolution and limit cycles of the 
    forced condensate, with driving amplitude $\Delta$ and frequency 
    $\Omega$ as indicated above the panels. The leftmost plots display
    a time span of eight periods of the driving force.}
\end{figure}

Figure \ref{fig:psplots} shows examples of the time evolution of 
$R(t)$, the solution of Eq.\ (\ref{req}), for two instances of 
the values of the parameters $\Omega$ 
and $\Delta$. 
The limit cycle, i.~e.\ the steady orbit that is 
approached at long times, in the phase space spanned by $R$ 
and $\dot{R}$ is also shown.
The damping parameter is fixed at $\gamma=0.02$.
In the first example, the forcing frequency $\Omega=2.2$ is comparable 
to the quadrupole frequency and the forcing amplitude is rather low, 
and therefore the motion is in this case nearly 
sinusoidal (so that the limit cycle takes on a nearly elliptic shape)
and locked to the forcing frequency. For a larger amplitude $\Delta$ 
the shape of the oscillations is distorted due to the 
strongly anharmonic effective potential well that $R$ resides in.
When the forcing frequency is smaller than an integer fraction of the 
breathing-mode frequency, $\Omega<\sqrt{5}/n$, an $n$-th harmonic is 
seen if $\Delta$ is large enough 
(but the motion is still periodic with period 
$\Omega$). This is in Fig.\ \ref{fig:psplots} 
illustrated for $\Omega = 0.53 = \sqrt{5}/4.2$ where four loops are 
seen in the limit cycle. These variational results have been compared to 
the solution of the full GPE (\ref{gpe}), and are in excellent 
agreement.

\begin{figure}
    \includegraphics[width=0.9\columnwidth]{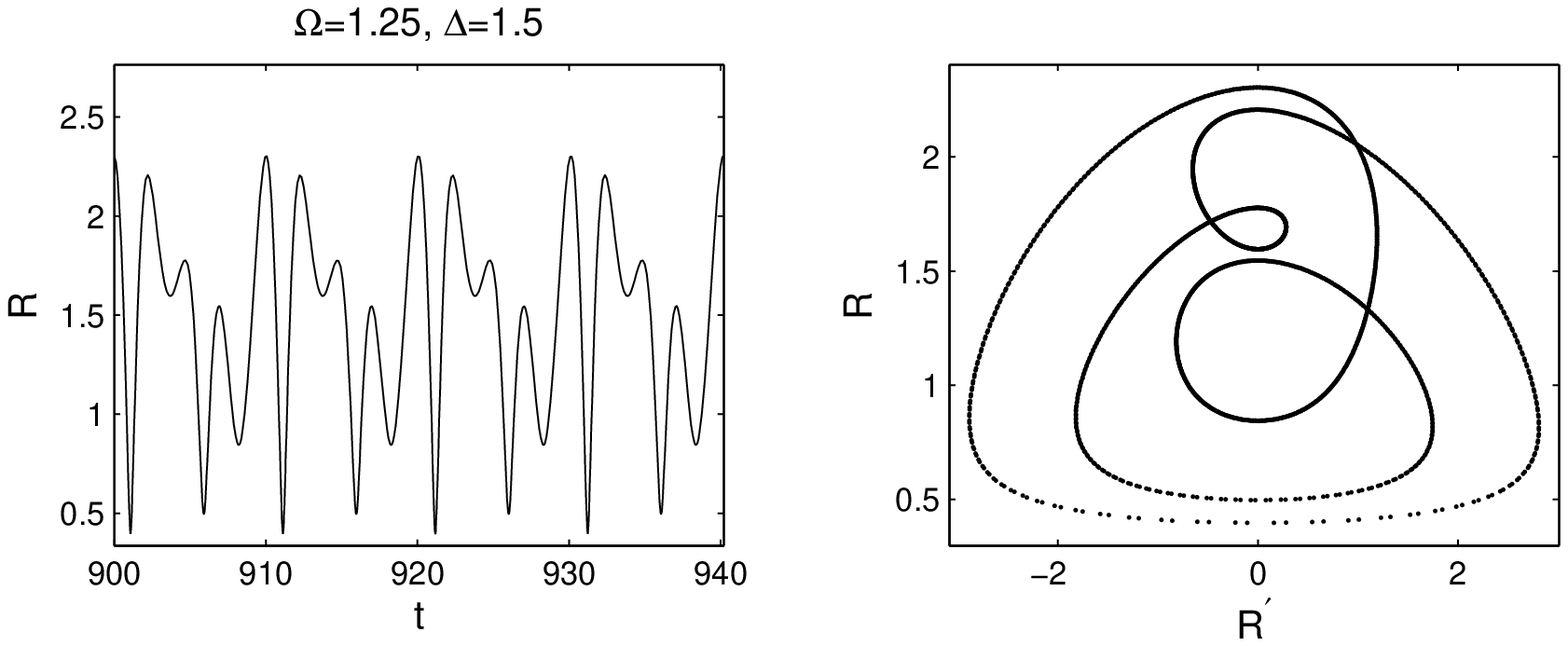}
    \includegraphics[width=0.9\columnwidth]{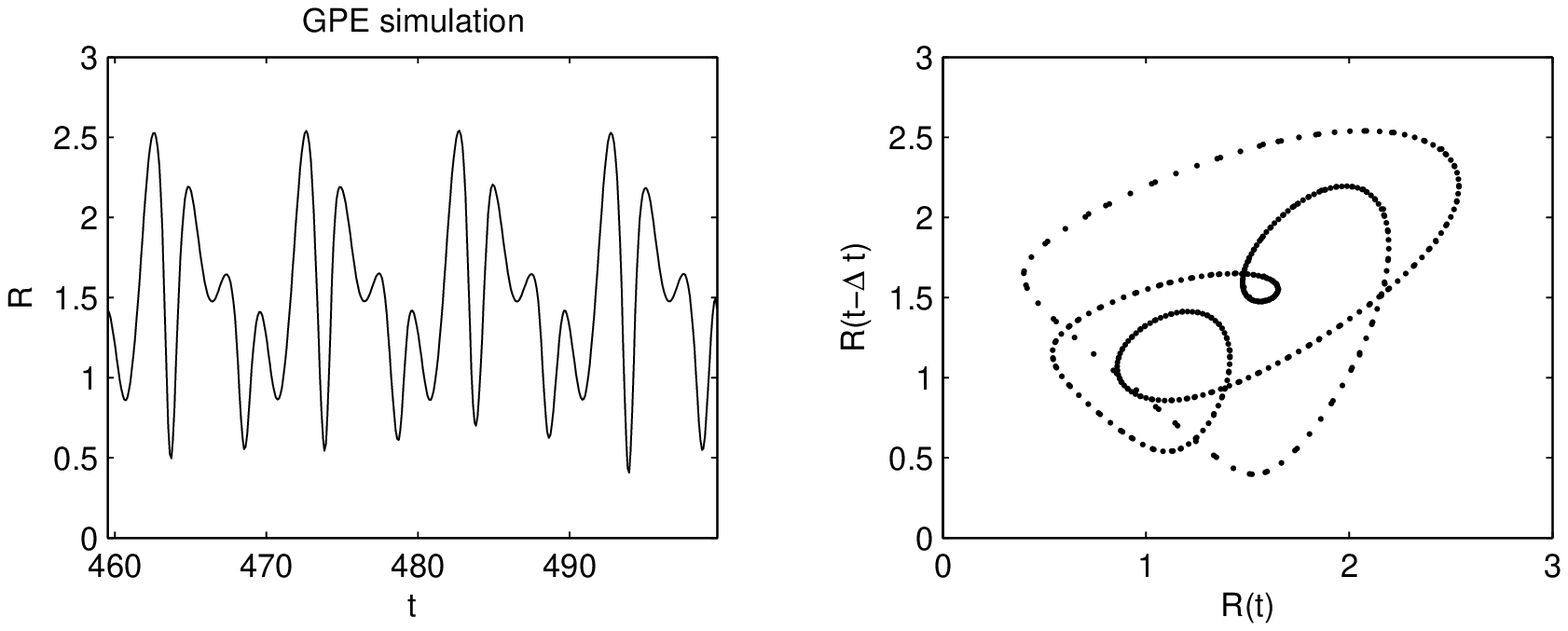}
    \caption{\label{fig:bifurc} Time evolution and limit cycles of the 
    forced condensate. Variational results are shown in the upper 
    panels, while the lower panels show the results of the full 
    Gross-Pitaevskii equation.The delay parameter for the limit cycle 
    in the lower plot is chosen as $\Delta t=0.5$.}
\end{figure}

\begin{figure}
    \includegraphics[width=\columnwidth]{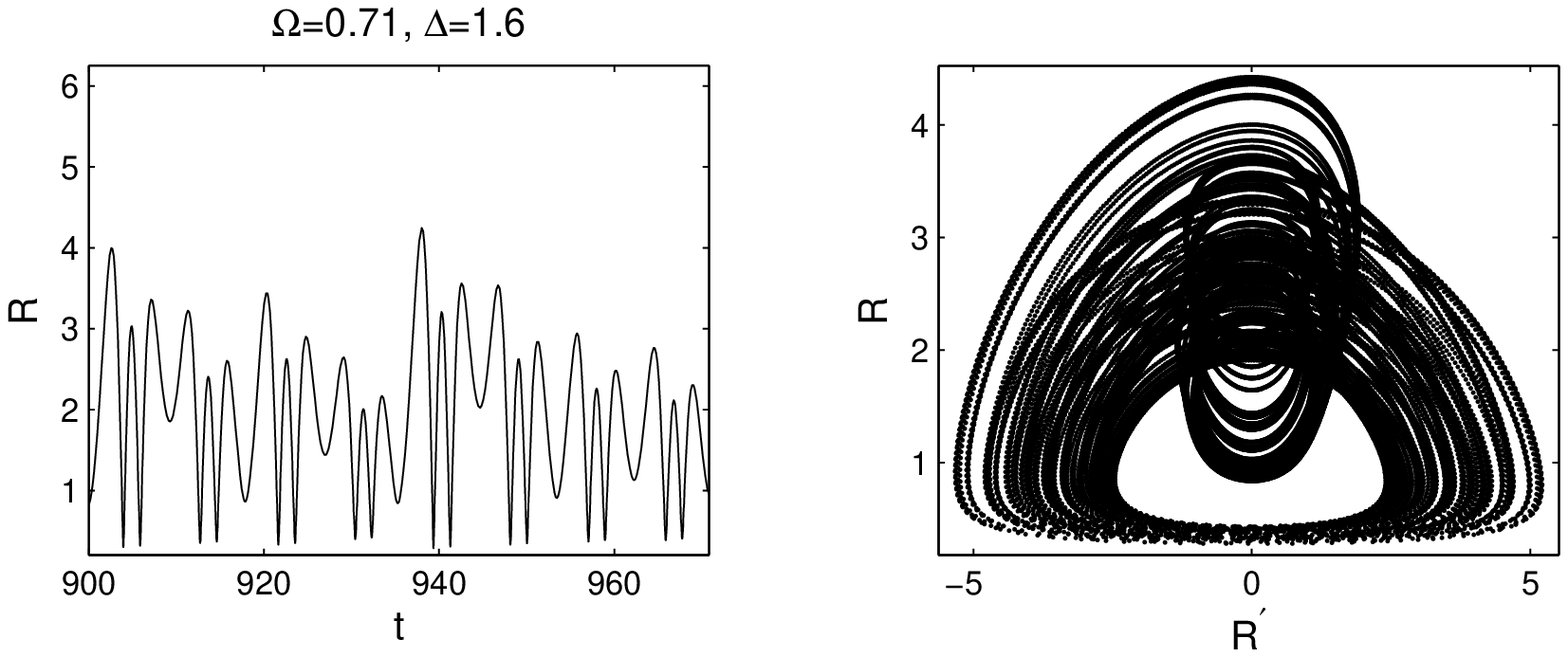}
    \includegraphics[width=\columnwidth]{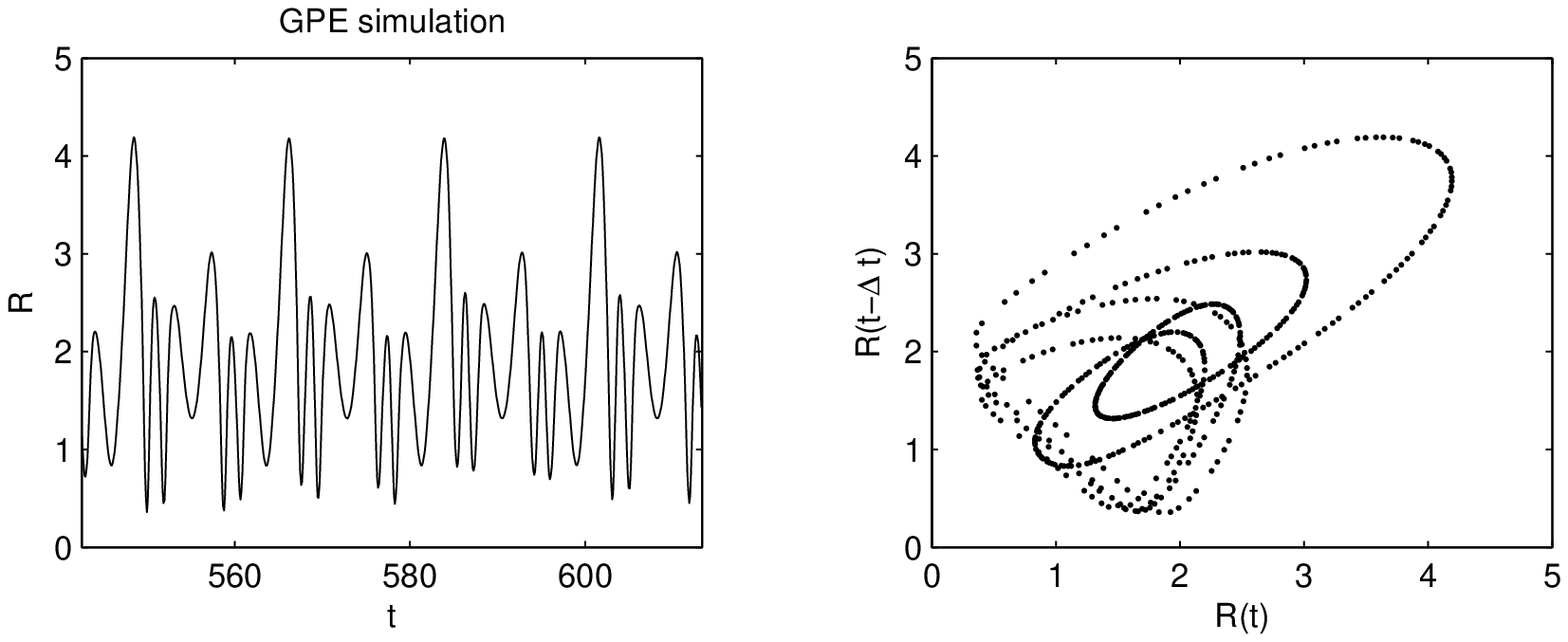}
    \caption{\label{fig:chaos} Time evolution and limit cycles of the 
    forced condensate. Variational results are shown in the upper 
    panels, while the lower panels show the results of the full 
    Gross-Pitaevskii equation.}
\end{figure}

For larger values of the forcing amplitude $\Delta$ more complex 
behavior is observed. 
When $\Omega$ is close to resonance with the breathing-mode 
frequency $\Omega=\sqrt{5}$ or fractions 
thereof \cite{fotnot2}, 
and $\Delta$ is large 
enough, the variational system 
(\ref{req}) exercises period-doubling bifurcations and chaotic motion. 
Figure \ref{fig:bifurc} 
shows the period-$2$ motion seen 
at $\Omega=1.25 \approx \sqrt{5}/2$, $\gamma=0.02$ and $\Delta=1.5$.
Shown are both the variational result for $R$ and 
the rms radius, $\sqrt{\langle r^2\rangle}$, of the 
cloud as calculated using the Gross-Pitaevskii equation, scaled by the 
Thomas-Fermi value $R_{rms}=\sqrt{3/7}\rtf$ to make the scale agree with 
that of the variational study. The limit cycle is plotted in the 
space of $R(t)$ and $R(t-\Delta t)$, where $\Delta t$ is a small time
delay; this yields an image tilted by 45 degrees compared 
to that in $R$-$\dot{R}$ space.
The variational model still agrees well with the full GPE. However, 
concerning chaos the situation is different: the solutions to 
the full GPE fail to exhibit chaotic motion when the variational 
does. An example is shown in Fig.\ \ref{fig:chaos}, 
where the parameters $\Omega=0.7$, $\Delta=1.8$ and 
$\gamma=0.02$ are chosen. 
That the solution to the effective-radius equation is indeed chaotic 
has been checked by calculating the Lyapunov 
exponent. 
However, the solution to the full Gross-Pitaevskii equation is 
seen to be perfectly regular: still period-doubled, 
albeit with a rather large amplitude. 
The grid parameters have been chosen so that the wave function both 
in configuration and momentum space stays well inside the grid 
boundaries at all times; grids up to 1500 points with a grid 
constant $\Delta x=0.04\aos$ have been tried.

\begin{figure}
    \includegraphics[width=\columnwidth]{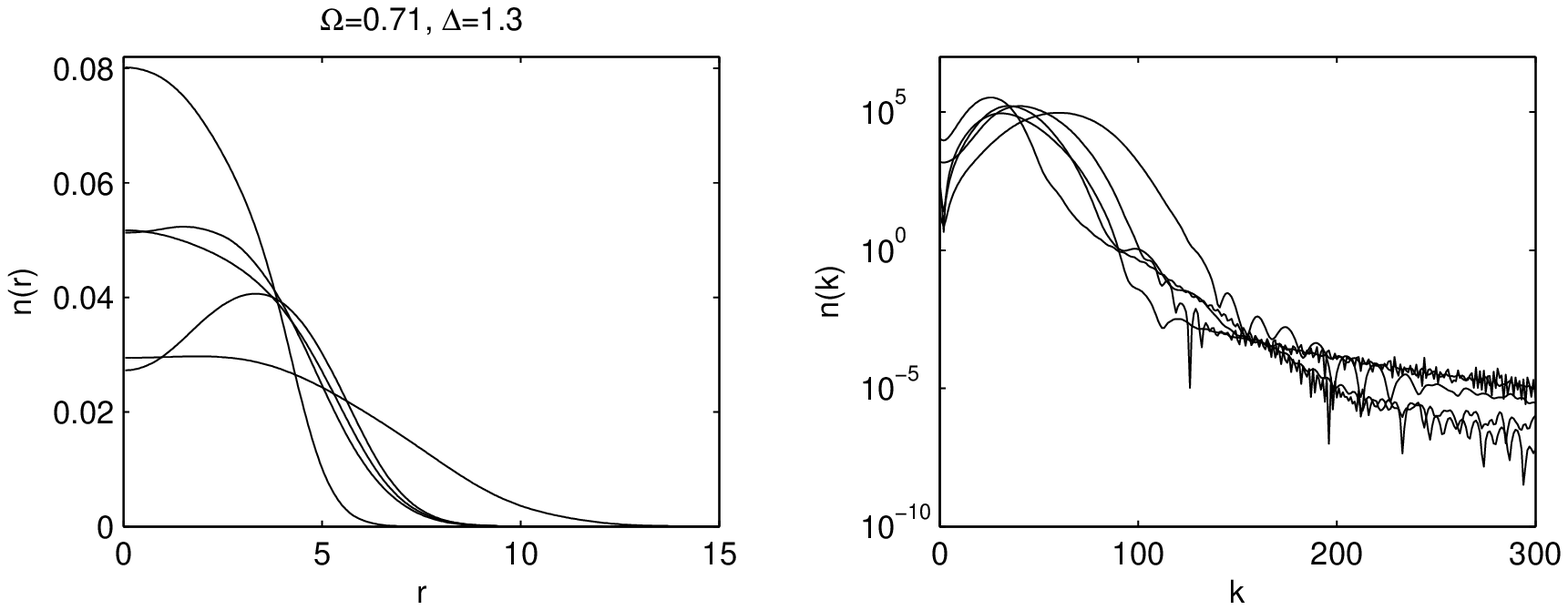}
    \includegraphics[width=\columnwidth]{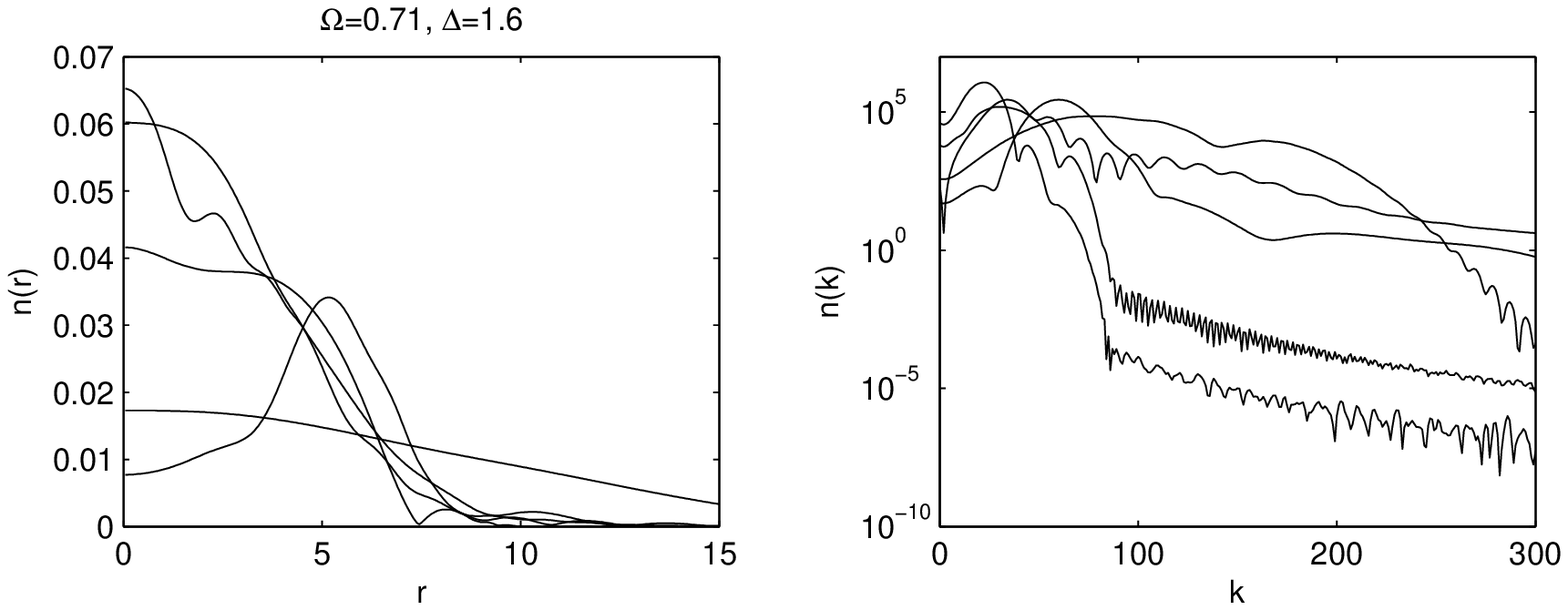}
    \caption{\label{fig:wf} 
      Density distribution at a few time instances, as calculated
      by the full Gross-Pitaevskii equation. Radial real-space 
      density distributions are shown to the left and momentum-space 
      distributions to the right.
      Topmost panels: periodic motion with $\Omega=0.71$ and 
      $\Delta=1.3$. Lower panels: period-two motion with 
      $\Omega=0.71$ and $\Delta=1.6$; this case was
      predicted by the variational equation to be chaotic. 
      The different curves 
      in each panel are chosen at arbitrary equally spaced instances,
      but the same set of time instances were used in all panels.
    }
\end{figure}

The result is quite general: nowhere in phase space does chaotic 
motion, or even periodic motion with period longer than 
$2$, occur in the GPE solution.
The discrepancy is not due to the omission of the dispersion-related 
kinetic-energy term in Eq.\ (\ref{req}); that is readily checked by 
including such a term whereby the solution still exhibits chaotic 
motion. Rather, the fallacy of the variational Eq.\ (\ref{req}) is 
that it does not let the wave function change its shape during the 
dramatic changes of the trap potential between confining and 
repelling. This may come as a surprise, because a variational study is 
almost always seen to faithfully mimic the physics at least qualitatively 
\cite{cd1}. A closer look at the wave function gives an explanation. 
Figure \ref{fig:wf} shows a few snapshots of the wave function for the 
cases of period-one and period-two motion. 
In the regime where there is agreement between the variational 
equation and the GPE, the profile of the wave function does not change 
drastically, in accordance with the assumptions behind the variational 
equation, 
but in the regime predicted to be chaotic, waves propagate 
back and forth through the cloud. Clearly the variational system 
has too few degrees of freedom to faithfully model the behavior of 
the cloud in this regime. In a manner of speaking, the instability 
that manifests itself in short-wavelength density waves in the full 
GPE study, must instead find its outlet as chaotic motion in the 
restricted variational study.

\begin{figure}
    \includegraphics[width=\columnwidth]{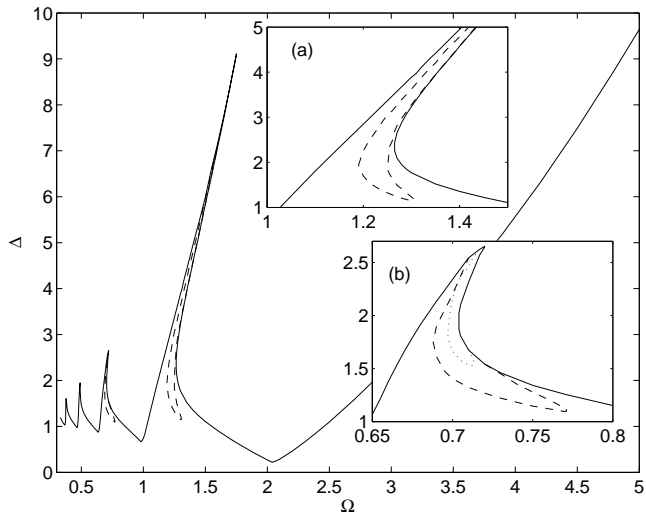}
    \caption{\label{fig:phasediag} Phase diagram for the forced 
    condensate. Solid lines enclose the regime of stable motion;
    dashed lines enclose the regimes of period-doubled  motion
    and the dotted line encloses the regime where the 
    variational study displays chaotic motion. The insets 
    labeled (a) and (b) zoom in on two important regions.}
\end{figure}

We conclude by showing in Fig.\ \ref{fig:phasediag} 
the full phase diagram of the forced 
system as calculated by the variational equation. The damping 
parameter is still $\gamma=0.02$.
Small regions of 
period-multiplied motion and chaos are seen to exist around the 
stability tongues just above $\Omega=\sqrt{5}/n$
for all integers $n>2$. The largest chaotic regime is also 
indicated although it was found to be an artifact of the 
variational equation.

To summarize, the Gross-Pitaevskii equation in a trap with a 
periodic modulation of the trapping potential has been seen to 
support period-doubled oscillations, but chaotic motion was not 
observed. In contrast, a simplified description of the same 
system using a collective coordinate predicts chaotic motion 
and period-multiplied motion with period higher than two, 
because of its failure to describe short-wavelength density 
modulations.
Indeed, it was suggested in 
Refs.\ \cite{GarciaLong,CastinInstability} that chaotic motion cannot 
occur in Bose-Einstein condensed systems; however, it seems that the 
reasoning leading to that conclusion is not applicable in the present case. 
Reference \cite{CastinInstability} 
proposed that the exponentially growing separation between 
neighbouring orbits in phase space that is characteristic of chaos 
implies that the depletion of the condensate grows equally fast, and 
Ref.\ \cite{GarciaLong} suggested that this can be taken as proof that 
chaotic oscillation in Bose-Einstein condensates is impossible. 
However, said analysis does not address 
the case of bounded chaotic oscillations where the exponential separation 
of nearby orbits only takes place over short times while at long times the 
whole bundle of orbits stays bounded, as in Fig.\ 
\ref{fig:chaos}. Furthermore, if the excited modes are coherent with 
the condensate, it is not at all clear that their growth must be 
identified with depletion of the condensate. 
The absence of chaos observed here should therefore be seen as an 
accidental fact, and
the question of whether 
chaotic oscillations in fact can be observed 
for slightly different physical situations -- for example exploring 
other parameter values or driving 
modes with higher multipolarity -- still remains open.

Discussions with Antti Kupiainen, Chris Pethick, Kalle-Antti 
Suominen, Matt Mackie 
and Simon Gardiner are gratefully acknowledged. 
This work was supported by the G{\"o}ran Gustafsson Foundation,
the Academy of Finland (Project 50314), 
and by the European Network ``Cold atoms 
and Ultra-Precise Atomic Clocks'' (CAUAC).


\begin{thebibliography}{99}
\bibitem{fotnot1}{Remember that a fourth parameter, the 
  coupling strength $Na/\aos$, is here assumed to be very large, whereby 
  it dropped out of the problem.}
\bibitem{fotnot2}{Since the breathing-mode frequency $\omega\approx 2.23$ 
is close to twice the bare trap frequency, it is impossible to 
tell from the presented numerical results alone whether it is the bare 
or the breathing-mode frequency that sets the scale for the onset of 
period doubling. 
However, with a general exponent $n$ in the inverse power law the 
natural frequency of the system is $\sqrt{n+1}$; trial 
calculations indicate that increasing $n$ 
accordingly leads to an increase of the frequencies for the chaotic 
regions.}
\end{thebibliography}
\end{document}